# TRGCN: A Hybrid Framework for Social Network Rumor Detection

Yanqin Yan[12], Suiyu Zhang[12], Dingguo Yu[12]*,Yijie Zhou[12],Cheng-Jun Wang[3]*,Ke-ke Shang[3]*

[1] School of Media Engineering, Communication University of Zhejiang, Hangzhou, 310018, China

[2] Zhejiang Key Laboratory of Film and TV Media Technology, Hangzhou, 310018, China

[3] Computational Communication Collaboratory, Nanjing University, Nanjing 210023, China

* Corresponding author: Dingguo Yu, Email: yudg@cuz.edu.cn; Cheng-Jun Wang, wangchengjun@nju.edu.cn; Ke-ke Shang, kekeshang@nju.edu.cn.

## Abstract

Accurate and efficient rumor detection is critical for information governance, particularly in the context of the rapid spread of misinformation on social networks. Traditional rumor detection relied primarily on manual analysis. With the continuous advancement of technology, machine learning and deep learning approaches for rumor identification have gradually emerged and gained prominence. However, previous approaches often struggle to simultaneously capture both the sequential and the global structural relationships among topological nodes within a social network. To tackle this issue, we introduce a hybrid model for detecting rumors that integrates a Graph Convolutional Network (GCN) with a Transformer architecture, aiming to leverage the complementary strengths of structural and semantic feature extraction. Positional encoding helps preserve the sequential order of these nodes within the propagation structure. The use of Multi-head attention mechanisms enables the model to capture features across diverse representational subspaces, thereby enhancing both the richness and depth of text comprehension. This integration allows the framework to concurrently identify the key propagation network of rumors, the textual content, the long-range dependencies, and the sequence among propagation nodes. Experimental evaluations on publicly available datasets, including Twitter 15 and Twitter 16, demonstrate that our proposed fusion model significantly outperforms both standalone models and existing mainstream methods in terms of accuracy. These results validate the effectiveness and superiority of our approach for the rumor detection task.

**Introduction**

As human societies have evolved, rumors have exhibited cascading dynamics within complex social systems, propagating through interconnected networks and often triggering nonlinear informational diffusion processes (Adrien et al., 2014). Over the last several years, the swift growth of the internet, the widespread proliferation of social media, and the increased portability of modern media devices have greatly accelerated the dissemination of information. Everyone can now publish, comment, and share information through personal social accounts. Moreover, the rise of AI-generated content (AIGC) has contributed to an overwhelming increase in false information, leading to exponential growth in rumors (Sun et al., 2024).

A rumor is a broadly disseminated piece of information with uncertain veracity, that seems plausible at first but ultimately confusing and difficult to distinguish from the truth (Barsha et al., 2023; Priyanka et al., 2020). The circulation of rumors may bring about severe adverse effects on both people and communities, influencing everything from personal lives to group behavior, and even disrupting public order and political activities. For instance, the spread of false information about the side effects and efficacy of the COVID-19 vaccine on social media fueled fear and resistance, significantly hindering vaccination efforts (Wang et al., 2021). To create a clear cyberspace for the public and combat rumors on social media, fact-checking platforms like Snopes.com and PolitiFact.com depend on manual verification by human experts to identify and refute rumors. However, these approaches have clear limitations in terms of efficiency and coverage. Therefore, it is essential to leverage new technological tools to improve the effectiveness of rumor identification. Existing methods for automatic rumor detection may be divided into feature-oriented approaches and deep learning-driven techniques.

Early studies (Castillo et al., 2011; Kwon et al., 2013; Liu et al., 2015) relied on manual feature engineering to train rumor classifiers. For example, Yang et al. (2012) extracted 19 handcrafted features from Weibo posts, such as reply frequency and repost counts, and trained an RBF-kernel SVM for classification. Our previous work (Zhang et al., 2022) also involved the manual extraction of features related to the content attributes and propagation structures of information cascades to predict news credibility. Nevertheless, the effectiveness of these features was found to be relatively limited. However, the higher-dimensional feature representations in rumor propagation cannot be captured by those methods.

In contrast, deep neural network-based methods (Ma et al, 2016; Jin et al., 2017) were employed to model temporal dynamics in rumor propagation; the LSTM and GRU models marked the first application of deep learning in capturing sequential rumor representations. Ma et al. (2016) proposed a Recurrent Neural Network architecture for detecting temporal sequences in social media posts, while Yu et al. (2017) employed convolutional neural networks (CNNs) for high-level textual representations to rumor detection. Nevertheless, the effectiveness of these features was found to be relatively limited, due to the inability of these methods to capture the high-dimensional and complex patterns in rumor propagation.

For comprehensive representation, Graph Neural Networks (GNNs) (Dou et al., 2021; Dong et al., 2019) were introduced, given their effectiveness in social network analysis. Wu et al. (2015) applied an undirected graph to encode global propagation structures. Bian et al. (2020) proposed a Bi-GCN model that leverages both top-down and bottom-up propagation patterns for dual-feature analysis. Graph Convolutional Networks (GCNs) are well-suited for processing graph-structured data and have demonstrated effectiveness in identifying potential rumor propagation patterns grounded in the architecture of social networks. As shown in **Fig. 1**, the propagation graph of a message randomly taken from Twitter15 reveals that even at the fifth level of propagation, nodes could still serve as primary hubs for further dissemination. Capturing long-range dependencies and the sequential order among nodes plays an important role in propagation graph feature extraction.

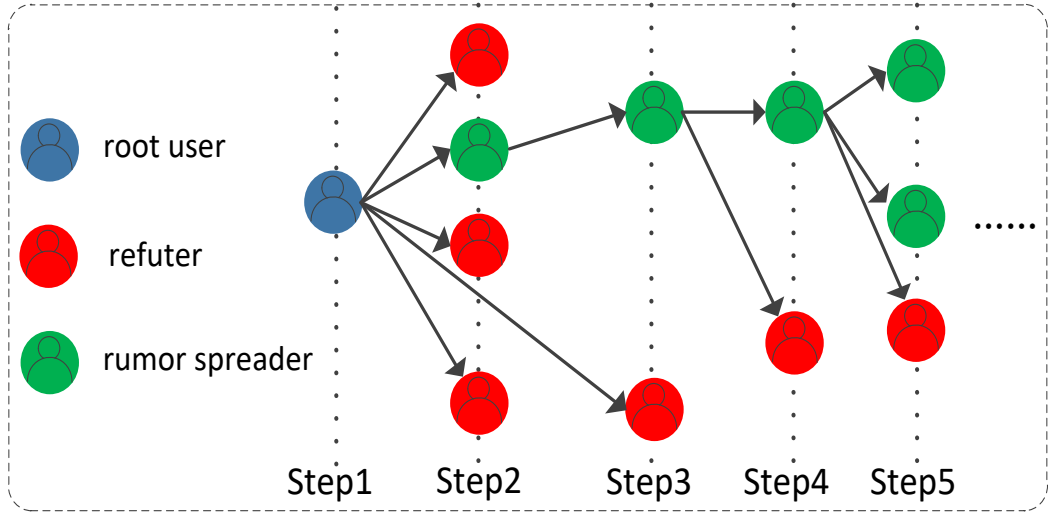

**Fig.1 Example of a rumor propagation network.** Each node in the network represents an individual social media user who interacts with a particular post containing the rumor. The blue node indicates the root user, the original poster who first introduces the piece of information (often the source of the rumor). Red nodes represent refuters, users who actively respond to or challenge the rumor by posting corrections, counterarguments, or factual information aimed at debunking the false claim. Green nodes are rumor spreaders, individuals who share or forward the rumor without questioning its validity, thus contributing to its dissemination. The diagram illustrates how the rumor spreads among different types of users over time steps. Arrows denote the forwarding relationships between users and the post.

However, previous research has highlighted that a key limitation of GCNs is their inability to effectively capture long-range dependencies between users. Xu et al. (2018) demonstrated that conventional GNNs have difficulty capturing key information from k-hop neighbors, while Li et al. (2018) noticed a substantial decline in the performance of GNNs with an elevated depth in neighborhood aggregation. Furthermore, Chen and Wong (2020) specifically identified the inadequacy of GNN-based sequence recommendation methods in modeling long-range dependencies, noting that few GNN layers restrict the model's ability to effectively learn relationships between distant items within sessions.

The Transformer was introduced by Vaswani et al. (2017) as a self-attention-based sequence-to-sequence model. The architecture efficiently captures long-range dependencies and sequential patterns in data. Thanks to its powerful feature extraction capability, the Transformer model has enabled several pivotal advancements in natural language processing, including but not limited to, sentence vector characterization through bi-directional encoder representations (Devlin et al., 2019), and dialogue system modeling (Tao et al., 2018).

To address these challenges, we introduce a hybrid model TRGCN for social network rumor detection, which performs multi-level fusion of GCN and Transformer. The GCN effectively captures local structural information within the graph, such as the relationships between nodes and their neighbors, and extracts relevant features from the graph's topology. The Transformer encoder leverages its self-attention for global context modeling, and models relationships between nodes at greater distances. This empowers the model to assess and integrate attributes in a way that produces richer and more comprehensive node representations. Moreover, the Transformer's Multi-Headed Attention Component handles dense connections across the graph, preserving the distinct characteristics of individual nodes. This enhances the model's expressive capacity, enabling it to retain nuanced differences between nodes. Additionally, the parallel computing capabilities of the Transformer significantly improve training efficiency and performance, particularly for large-scale graph processing. Our contributions are:

1. We propose an innovative framework that merges the GCN and Transformer models to achieve complementary advantages, leveraging the structural and textual details within social network data to identify rumors more thoroughly and precisely, and to enhance the accuracy and efficiency of the rumor detection model.

2. The Multi-level structure is designed to: (1) comprehensively and meticulously analyze the content complexity of rumor texts; (2) analyze the rumor propagation structure in multiple dimensions; (3) better utilize contextual information to assist in rumor judgment; (4) accurately capture hidden false information and misleading clues; and (5) refine veracity assessment in rumor identification.

3. Evaluations of the public datasets Twitter15 and Twitter16 reveal the enhanced robustness and efficacy of our model, particularly in achieving superior accuracy during the quad classification for rumor identification.

This paper proceeds as follows: Section 2 explores related studies, laying the foundation for our study. Section 3 delineates the suggested model TRGCN, explaining its design and functionality. Section 4 presents and discusses the outcomes of experiments, examining the efficacy and implications of our model. Finally, Section 5 summarizes the key findings, concludes the study, and highlights the contributions as well as potential directions for future work.

**Related work**

With the extensive use of social media, rumors spread fast and have a wide range of influence, and accurate detection of rumors is essential to preserving the veracity of data and social stability. The research and development of rumor detection has been hotly debated over the past few years, and considerable advancements have been achieved.

Early rumor detection relied heavily on manually developed rules and statistical features. These approaches typically assess whether information qualifies as a rumor by analyzing lexical and grammatical features of the text, as well as basic characteristics of the publisher. For instance, Castillo et al. (2011) evaluated the credibility of information through a decision tree model that manually extracts features. Yang et al. (2012) and Liu et al. (2015) applied SVM models using manually engineered features. However, these methods often lack a deep understanding of textual semantics and complex network structures, resulting in limited accuracy in rumor detection.

With the advancement of deep learning, which is grounded in end-to-end learning, features can be automatically extracted without manual design, enabling models to better process high-dimensional data. Ma et al. (2016) employed RNNs for extracting latent representations of time-series feature data. Yu et al. (2017) introduced a CNN-driven technique that extracts dispersed input sequence characteristics and models salient feature interrelations. Ma et al. (2018) constructed propagation structures and hidden features of text through RvNN, improved efficiency, but did not take into account the high-dimensional and structural characteristics of information dissemination.

Huang et al. (2020) applied Graph Convolutional Networks for detecting rumors on social media and captured the relationships between nodes (e.g., users, posts) to enhance detection accuracy. Thota et al. (2023) leveraged Graph Convolutional Networks (GCNs) to model how rumors spread, pictured the source and subsequent responses as a branching tree by treating this propagation pattern as a graph, and constantly updated node representations to reflect the evolving context. Zhao et al. (2024) proposed an innovative framework for detecting rumors that employed a Graph Anchor Latent Dirichlet Allocation (GA-LDA) model to uncover the underlying thematic structure of rumor events. Following this, the framework leveraged Graph Attention Networks (GATs) to capture the intricate semantic relationships and propagation patterns inherent in these events. Lu & Li (2020) proposed a novel model called GCAN. The model employed convolutional and recursive neural networks for user-focused representation learning during the forward pass, leveraging user-specific attributes. They crafted graphs that mirror potential user interactions, subsequently harnessed the Graph Convolutional Network (GCN) to discern user interaction patterns with a graph-aware approach. Sun et al. (2022) created a unique approach, which creatively uses two sets of GCN networks to capture the structural patterns presented at different time points. However, it should be noted that GCN has certain limitations when it comes to handling long-distance dependencies and complex semantic information.

The Transformer architecture, first introduced for natural language processing (Vaswani et al., 2017), has since been widely adopted and has achieved significant success in computer vision applications (Shah et al., 2025). Fang et al. (2024) proposed a new Graph Neural Network architecture that took a unique approach in the GCN-Transformer framework, empowering the training process through meta-path information enhancement technology. This method aimed to improve the accuracy of predicting application (App) usage patterns. Tang et al. (2020) introduced a dual-Transformer architecture, which was improved by a dependency graph and could analyze flat representations derived from standard Transformers and graph structure representations based on dependency graphs in parallel during repeated interactions. On the other hand, research is also evolving towards the integration of sophisticated attention mechanisms. Huang et al. (2024) introduced an investigative strategy for discerning rumors by employing a Bidirectional Gated Recurrent Unit (Bi-GRU) model with Dual Self-Attention (DSA) integration. The Bi-GRU module is employed to effectively capture medium and long-term temporal dependencies within sequential data, while maintaining robust performance across variable-length inputs. The integrated DSA mechanism served to dynamically filter redundant information from the dataset, thereby optimizing feature representation and significantly enhancing the detection performance. Yu et al. (2022) proposed HDGCN, and Ye et al. (2023) introduced GCRES, both rely on the GCN module to capture structural information, lack mechanisms to model global long-range dependencies, and perform multi-level feature fusion across different modules. Their architectures were based on simple serial connections, resulting in a limited information flow and insufficient hierarchical feature extraction.

Therefore, based on a comprehensive study of rumor traits and the advancement of detection methods, we introduce an innovative rumor identification model that adds positional encoding at an early stage without compromising the original graph topology. This directly enriches node representations by assigning each node a unique identity. Through the multi-level fusion of Transformer and GCN, our model achieves high-precision modeling of both graph structures and sequential patterns, effectively capturing structural and textual features, thereby improving the accuracy and robustness of detection.

## Methodology

### *Preliminaries*

In this section, we define the problem formulation and introduce the key notations used throughout the paper.

The set of events for the rumor identification task is defined as $R = \{r_1, r_2, \cdots, r_n\}$, where $r_i$ represents the $i$-th rumor event and $n$ is the number of events. Each event $r_i = \{c_{i,0}, c_{i,1}, \cdots, c_{i,n-1}, G_i\}$ includes a source post (statement) and its associated comments, $c_{i,0}$ is the initial statement (root post) of the event $r_i$, $\{c_{i,1}, \cdots c_{i,n-1}\}$ represents the count of comments for the given statement, where $n$ equals the comment total for the event. For event $r_i$, its propagation process is modeled as a directed graph $G_i = (P_i, E_i)$, the node set $P_i = \{c_{i,0}, c_{i,1}, \cdots c_{i,n-1}\}$ denotes the statements and comments, and the edge set $E_i = \{e_{i,jk} | j, k = 0, \cdots, n-1\}$ consists of reply relations $c_{i,j} \rightarrow c_{i,k}$ denoting that a comment $c_{i,j}$ replies directly to $c_{i,k}$. According to the node and edge composition, get the neighbor matrix $A \in \{0,1\}^{N \times N}$. By integrating the structure of content and dissemination, we establish a neighbor matrix $Z$ for a heterogeneous graph.

Denote $M_i = [m_{i,0}^{\mathsf{T}}, m_{i,1}^{\mathsf{T}}, \cdots, m_{i,n-1}^{\mathsf{T}}]$ as a feature matrix extracted from the posts in $r_i$, where $m_{i,0}$ represents the feature vector of $c_{i,0}$ and each other row feature $m_{i,j}$ represents the feature vector of $c_{i,j}$.

For the rumor classification task, given the labeled Twitter rumor datasets $D = \{(r_i, y_i)\}$, where each $r_i$ corresponds to a fine-grained category label $y_i$. The objective of the exercise is to acquire a classifier:

$$f : R \rightarrow Y, Y \in \{nonrumor, falserumor, truerumor, unverifiedrumor\}. \qquad (1)$$

The dataset $D$ consists of multiple rumor events $r_i$, where each $r_i$ includes the source tweet and all reply tweets in its propagation tree $G_i$. Each event is associated with a label $y_i$, which is provided by expert annotations or authoritative fact-checking platforms (e.g., Snopes, PolitiFact).

### *TRGCN model Overview*

The proposed model integrates both the content of nodes and the structural characteristics of rumor propagation. It accounts for the sequential nature inherent in rumor dissemination and extracts features from multiple dimensions to facilitate rumor classification. This is achieved through the Multi-head attention mechanism of the Transformer architecture. As illustrated in **Fig. 2**, the model comprises three main modules, each of which will be fully explained in the next section.

The initial stage involves preprocessing information obtained from social media, with this information categorized into two components: content and propagation structure. Each component undergoes distinct feature representation techniques. Specifically, the analysis employs Term Frequency-Inverse Document Frequency (TF-IDF) (Ramos, 2003) for capturing textual relevance, while the propagation structure is represented through an adjacency matrix that reflects the network of interactions. These two representations are then integrated into a unified feature space. After that, we assign a positional encoding to each node to capture its relative location within the propagation graph. Each node's positional encoding is determined by its depth in the propagation tree: nodes closer to the root are assigned smaller indices, while deeper replies receive larger indices. For batch processing, positional indices are independently assigned within each graph, starting from zero. These indices are then mapped to positional encoding vectors using a combination of sine and cosine functions, following the standard Transformer approach. The resulting positional encoding vectors are integrated with the node features via element-wise addition, enriching each node representation with both textual information and structural position. This design allows the Transformer's attention mechanism to effectively incorporate the hierarchical structure of the propagation.

The second and central component of the model is the Multi-level processing framework. The model architecture employs a hierarchical processing pipeline that seamlessly transitions between graph-structured and sequence-structured representations. Initially, node features are processed through a Graph Convolutional Network (GCN) layer to capture localized structural dependencies and neighborhood information. Following this graph-level processing, the node embeddings are systematically transformed into sequence format through a structured padding and batching mechanism, where nodes from each graph instance are organized into fixed-length sequences while preserving their original graph membership. These sequence-formatted representations are then fed into a multi-layer Transformer encoder, which leverages attention mechanisms to model global dependencies and long-range interactions across the entire node sequence. Subsequently, the enriched sequence representations are meticulously reconstructed back into their original graph structure through a reverse mapping process, ensuring the preservation of graph-level integrity while incorporating the globally-aware contextual information derived from the Transformer processing. This sophisticated graph-sequence-graph transformation paradigm enables the model to effectively combine the complementary strengths of both graph convolutional operations for local structural modeling and self-attention mechanisms for global relational reasoning, thereby facilitating comprehensive representation learning.

The third element in the model constitutes the output stage, which is responsible for the final prediction of rumor classifications. Specifically, it performs a nonlinear transformation and dimension mapping on the information features extracted by the previous module through a fully connected layer, and then connects the softmax activation function to generate probability distributions of various categories, thereby achieving efficient discrimination and precise classification of rumors, and providing decision output for the rumor detection task.

Next, we elaborate on the specific mechanisms and implementations of each component.

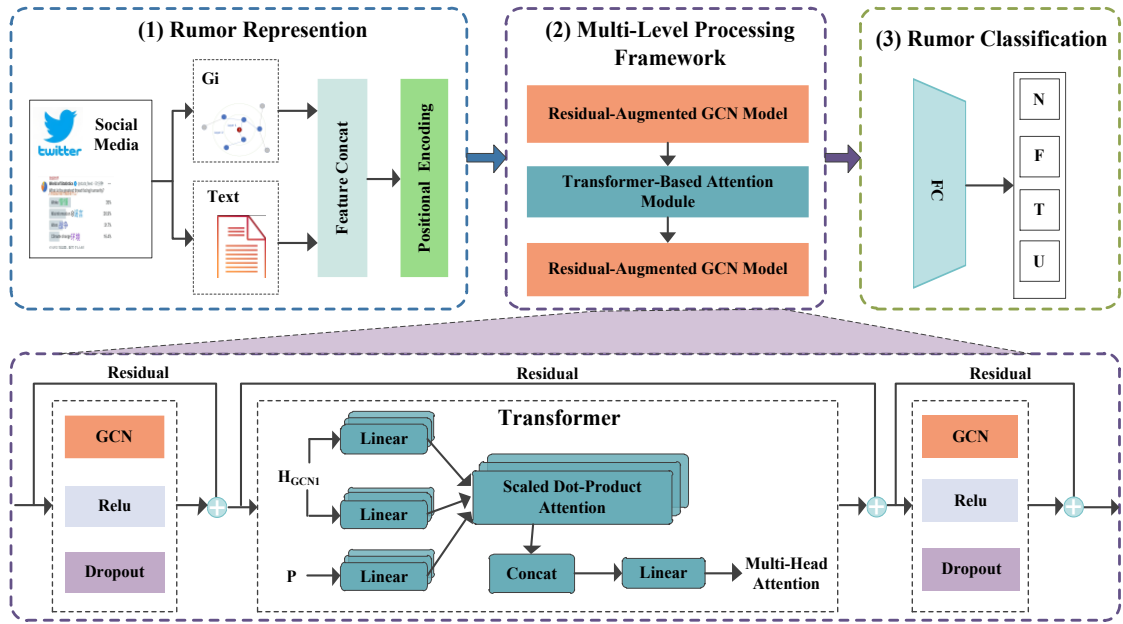

**Fig.2 The architecture of the TRGCN module.** TRGCN consists of three components: (1) rumor representation through the integration of textual content and propagation structure, followed by the integration of node features with positional encoding; (2) a multi-level fusion model that combines Graph Convolutional Networks (GCNs) and Transformer architectures, with their representations integrated through attention mechanism layers; and (3) the rumor classification task, which determines the veracity of a given piece of information (e.g., true rumor, false rumor, unverified rumor, or non-rumor).

***Rumor Representation***

To obtain a heterogeneous feature representation of rumors, we first represent the content and the propagation structure separately using appropriate encoding methods. These distinct representations were then fused to capture both semantic and structural aspects of the rumor data. Additionally, positional encoding was applied to the graph nodes to model the sequential characteristics of rumor propagation. This approach enables the model to effectively integrate diverse features and preserve the depth inherent in the propagation process.

***Feature Concat.*** To construct the content representation of a rumor post, we begin by calculating the TF-IDF score for each word. Term frequency $tf_{ij}$ is defined as $tf_{ij} = \dfrac{n_{ij}}{\sum_k n_{ik}}$ where $n_{ij}$ indicates the frequency of the word occurrence within the post $j$, and the denominator sums the occurrences of the word $i$ across all posts. The inverse document frequency $idf_i$ is defined as:

$$idf_i = \log(\frac{|D|}{1 + |\{k : t_i \in p_k\}|}). \tag{2}$$

Given that $|D|$ it is the corpus post count, and $|\{k : t_i \in p_k\}|$ denotes posts containing a word $t_i$. The weight of the word $t_i$ is computed as $t_i = tf_{ij} \times idf_i$. The content of the post $j$ is represented as $p_j = [t_1, t_2, \cdots, t_{|W|}]$, where $|W|$ is the number of corpora. Given that the representation of $p_j$ is sparse and high-dimensional, we use an embedding layer to reduce its dimension. Formally, the dense vector of each rumor content is defined as:

$$m_j = W_j p_j, \tag{3}$$

where $W_j$ denotes the embedding layer's weight matrix. $m_j$ is the content representation of post $j$.

We define each event as consisting of content and its propagation structure. Based on the constructed adjacency matrix $A \in \{0,1\}^{N \times N}$, we obtain a new adjacency matrix $Z$, which is defined as:

$$Z = A \cdot m_j, \tag{4}$$

where $Z \in R^{N \times d}$ not only retains the original content's semantic meaning but also incorporates the associative relationships between the nodes, enabling the features of each node to perceive its contextual information in the graph structure.

***Position Encoding.*** To enable the model to distinguish based on node positions in the graph, it is essential to incorporate positional encoding into the node representations. Since graph structures lack inherent order, unlike sequences, positional encoding provides a mechanism for capturing the positioning, either relative or exact, of network nodes. This enhances the model's ability to recognize structural roles and spatial relationships among nodes, which is critical for accurately modeling information propagation in the network.

Positional encoding constitutes a fundamental element of the Transformer architecture. This compensates for the lack of sequential awareness of the attention mechanism by explicitly injecting position information. The Transformer employs both sine and cosine functions to generate positional encodings, as defined by the following formula:

$$PE_{(pos,2i)} = \sin\left(\frac{pos}{10000^{\frac{2i}{d_{model}}}}\right), \tag{5}$$

$$PE_{(pos,2i+1)} = \cos\left(\frac{pos}{10000^{\frac{2i+1}{d_{model}}}}\right). \tag{6}$$

Here, pos denotes the position of a token in the sequence (ranging from 0 to $L-1$), where $L$ is the sequence length. The index $i$ represents the dimension index in the embedding vector, $d_{model}$ denotes the dimensionality of the word embeddings.

The input sequence is $Z \in R^{N \times d}$ with $\mathrm{N}$ denoting sequence length and $d$ signifying characteristic space. Position encoding matrix $PE \in R^{L \times d}$, where $L$ is a predefined maximum sequence length satisfying $L \geq N$. First, a slicing operation is performed on the position encoding matrix $PE$, extracting the first $N$ rows to obtain $PE_{slice} \in R^{N \times d}$,

$$PE_{slice} = \begin{bmatrix} PE_{1,1} & PE_{1,2} & \cdots & PE_{1,d} \\ PE_{2,1} & PE_{2,2} & \cdots & PE_{2,d} \\ \vdots & \vdots & \ddots & \vdots \\ PE_{N,1} & PE_{N,2} & \cdots & PE_{N,d} \end{bmatrix}. \tag{7}$$

The sliced positional encoding matrix $PE_{slice}$ is then combined with the input sequence $Z$ to form the output sequence with position information $Z' \in R^{N \times d}$, which is mathematically expressed as:

$$Z'_{i,j} = Z_{i,j} + PE_{slice_{i,j}} \qquad for \, i = 1, \cdots, N; j = 1, \cdots, d \tag{8}$$

Here, $Z'_{i,j}$ represents the element at position (i,j) in the output sequence $Z'$, $Z_{i,j}$ the corresponding element in the input sequence $Z$, $PE_{slice_{i,j}}$ denotes the element at position $(i,j)$ in the sliced position encoding matrix $PE_{slice}$.

### Multi-level Processing Framework

**Residual-Augmented GCN Model.** The model receives two types of heterogeneous inputs that encode social network structure and textual content features. The social relationship graph captures the interactions (e.g., follows, replies, retweets) among nodes (users/posts) via an adjacency matrix $A \in \{0,1\}^{N \times N}$, where N denotes the overall count of nodes. After initialization and regularization of the attributes of the nodes, each node $V_i$ is associated with an initial feature vector $p_i \in R^{d_0}$, which is usually randomly initialized or a pre-trained shallow embedding. Add a self-loop and symmetrically normalize the adjacency matrix:

$$\hat{A} = \widetilde{D}^{-\frac{1}{2}} \widetilde{A} \widetilde{D}^{-\frac{1}{2}}, \tag{9}$$

where $\widetilde{A} = A + I$ is the adjacency matrix with self-loops.

Let $A$ and $I$ stand for the adjacency and identity matrices of the heterogeneous graph, respectively, while $\widetilde{D}$ represents its degree matrix. From these components, we establish the initial state representation for the integrated graph. The model's governing equation takes the following form:

$$H_1 = \sigma(\hat{A} Z' W_0). \tag{10}$$

In our analysis, the initial stage of rumor dissemination is denoted as $H_1$, and the $W_0$ serves as a key parameter. We have opted $\sigma(\cdot)$ for the ReLU function to serve as our activation mechanism. Additionally, dropout techniques are integrated into the Graph Convolutional Network (GCN) layers to prevent overfitting. The residual architecture mitigates issues like gradient vanishing during the training of deep neural networks, thereby facilitating more stable and efficient optimization. The output equation is:

$$H_{GCN1} = Z' + H_1. \tag{11}$$

***Transformer-Based Attention Module.*** The Multi-head attention mechanism extracts features simultaneously from diverse subspaces, augmenting the model's proficiency in interpreting input sequences. Through the application of the Multi-head attention mechanism, we convert the present local hidden state into a global one. This procedure motivates us to construct a model to enhance the efficacy of rumor identification. We calculate the attention weight between the news node $V$ and the hidden state $H_{GCN1}$ to ascertain the relative significance in information dissemination. Mapping the input to $h$ subspaces parallelizes the computation of attention, and the Multi-head attention outputs are combined through concatenation to form the aggregated node representation $H_{mh}$. The formulas are as follows:

$$H_{mh} = MultiHead(Q, K, V) = Concat(head_1, \cdots, head_h)W^0. \tag{12}$$

Each attention head computes the following:

$$head_i = Attention(QW_i^Q, KW_i^K, VW_i^V).$$ (13)

As mentioned above, $P$ denotes the primary encoding of news nodes, while $H_{GCN1}$ denotes the hidden state. The scaled dot-product attention mechanism is defined as:

$$Attention(P, H_{GCN1}, H_{GCN1}) = soft\max\left(\frac{PH_{GCN1}^T}{\sqrt{d_k}}\right) \cdot H_{GCN1}.$$ (14)

We apply layer normalization with residual concatenation to stabilize the training process and preserve multilevel features.

$$H_{attn} = LayerNorm(H_{GNC1} + Dropout(H_{mh})).$$ (15)

Next, we apply a feed-forward network (FFN) to transform the attention output:

$$H_{trans} = LayerNorm(H_{attn} + Dropout(\operatorname{Re}LU((H_{attn}W_1 + b_1)W_2 + b_2))),$$ (16)

here, $W_1 \in R^{d \times d_{fin}}, W_2 \in R^{d_{fin} \times d}, d_{fin} = 4d$.

Before being fed into the classification model, the input is further processed by a residual GCN network, whose structure is the same as the initial residual GCN module, with inputs $H_{trans}$ and outputs $H_{final} \in R^{N \times d}$.

### *Rumor Classification*

Finally, we use a dense layer with softmax activation for rumor detection, defined as:

$$\hat{Y} = Soft\max(W_e H_{final} + b_e), \qquad W_e \in R^{4 \times 4},$$ (17)

where $W_e$ and $b_e$ are the weights and biases of the final hidden layer, and $\hat{Y} \in R^4$ signifies a probability vector across all classes, used to predict the classification of the misinformation event.

The model is trained by optimizing the cross-entropy loss between predicted $\hat{Y}$ and true distributions $Y$, with an $L_2$ regularizer incorporated into the loss function to mitigate overfitting.

## Experiment

We present the performance of our model on the public dataset and compare it with that of the benchmark model. Our method achieves higher accuracy in rumor classification.

### *Experimental datasets*

For validation, experiments employed open-access, actual social media data to evaluate our technique. These are Twitter15 and Twitter16 (Ma et al., 2017). Twitter15 contains 1490 propagation graphs, while Twitter16 has 818 propagation graphs. In all datasets, nodes denote posts, edges indicate retweet or reply connections, and features consist of the top 5000 TF-IDF weighted words. Each graph is annotated as one of four classes: non-rumor, true-rumor, false-rumor, and unverified-rumor. **Table 1** showcases the statistics of the two datasets.

**Table 1.**

*Statistics of the datasets*

| Statistic | Twitter15 | Twitter16 |
|:---:|:---:|:---:|
| * of posts | 331612 | 204820 |
| * of events | 1490 | 818 |
| * of non-rumors | 372 | 205 |
| * of true-rumors | 374 | 203 |
| * of false-rumors | 370 | 205 |
| * of unverified-rumors | 374 | 205 |

### *BaseLines*

To gauge the effectiveness of our TRGCN model, we pitted it against a range of established rumor detection techniques. These baselines spanned the gamut from early machine learning approaches and classic deep learning architectures to the more contemporary graph neural networks and their variants. We also evaluated the basic GCN-Transformer (GT-Base) model and a variant of our model in which the data preprocessing component was replaced with BERT representations, as shown in **Table 2**.

**DTC**: This approach uses a Decision Tree classifier with manually engineered features for tweet extraction.

**RFC**: This approach utilizes a Random Forest classifier incorporating user-based, linguistic, and structural news features for rumor identification.

**SVM-TS**: This approach implements a Support Vector Machine with a custom time-series kernel based on handcrafted features.

**SVM-HK**: This approach employs an SVM classifier combined with a graph kernel designed to compare rumor propagation patterns.

**GRU-RNN**: The approach employs recurrent neural networks equipped with GRU layers to identify shifts in contextual data within pertinent tweet sequences throughout time.

**BU-RvNN**: The approach uses bidirectional, tree-structured GRUs (operating in top-down and bottom-up directions) to model rumor diffusion trees for misinformation detection.

**Rumor2vec**: The approach uses a CNN-based architecture that jointly learns text and diffusion pattern representations for rumor detection.

**HDGCN: A heterogeneous GCN approach for dynamic rumor detection, incorporating a combined content-propagation module and an ODE-driven dynamic component.**

**GCRES: A deep residual graph convolutional network for modeling content and structural interactions in the propagation graph.**

**GT-Base: A baseline GCN-Transformer model that combines a single-layer GCN with a Transformer encoder.**

**Bert-TRGCN: A TRGCN variant using BERT for text feature extraction and operating on a tree-structured propagation graph for classification.**

**Table 2.**

*Overview of baseline models*

| Model | Dataset | Evaluation metrics | Reference |
|---|---|---|---|
| DTC | Twitter | Precision, Recall, F-Score | Castillo et al., 2011 |
| RFC | Twitter | Accuracy, Precision, Recall, F-Score | Kwon et al., 2017 |
| SVM-TS | Weibo, Twitter | Accuracy, Precision, Recall, F-Score | Ma et al., 2015 |
| SVM-HK | Weibo | Accuracy, Precision, Recall, F-Score | Wu et al., 2015 |
| GRU-RNN | Weibo | Precision, Recall, F-Score | Yang et al.,2012 |
| BU-RVNN | Twitter (15,16) | Accuracy, F-Score | Ma et al., 2017 |
| Rumor2vec | Weibo,Twitter(15,16) | Accuracy, Precision, Recall, F-Score | Tu et al., 2021 |
| HDGCN | Twitter (15,16) | Accuracy, F-Score | Yu et al., 2022 |
| GCRES | Twitter (15,16) | Accuracy, F-Score | Ye et al., 2023 |
| GT-Base | Twitter (15,16) | Accuracy, F-Score | Fang et al.,2024 |
| Bert-TRGCN | Twitter (15,16) | Accuracy, F-Score | Devlin et al.,2019 |

### Experiment Setup

We implemented the deep learning models using PyTorch and PyTorch Geometric, conducting all experiments on an NVIDIA RTX 3090 GPU. Accuracy was adopted as the primary evaluation metric, with the F1-score included as a supplementary measure to account for potential class imbalance in the datasets. The model was configured with a post embedding dimension of 256 and two GCN layers, while the Multi-head attention mechanism employed 4 heads and a latent space of dimensionality 128. A dropout rate of 0.2 was applied to enhance generalization. The model was trained for 100 epochs using a learning rate of 0.0005 and a batch size of 64. To prevent overfitting, early stopping was implemented when the validation performance showed no improvement over 10 consecutive epochs.

### *Experimental results and performance*

Rumor detection results on Twitter15 are shown in **Table 3,** and Twitter16 is shown in **Table 4** (N: NonRumor; F: FalseRumor; T: TrueRumor; U: UnverifiedRumor).

**Table 3.**

*The results of comparative experiments on the Twitter15 dataset*

| Method | Acc | NR F1 | FR F1 | TR F1 | UR F1 |
|---|---|---|---|---|---|
| DTC | 0.454 | 0.733 | 0.355 | 0.317 | 0.415 |
| RFC | 0.565 | 0.810 | 0.422 | 0.401 | 0.543 |
| SVM-TS | 0.544 | 0.796 | 0.472 | 0.404 | 0.483 |
| SVM-HK | 0.493 | 0.650 | 0.439 | 0.342 | 0.336 |
| GRU-RNN | 0.641 | 0.684 | 0.634 | 0.688 | 0.571 |
| BU-RVNN | 0.708 | 0.695 | 0.728 | 0.759 | 0.653 |
| Rumor2vec | 0.796 | 0.883 | 0.746 | 0.836 | 0.723 |
| HDGCN | 0.834 | 0.853 | 0.868 | 0.859 | 0.823 |
| GCRES | 0.853 | 0.855 | 0.858 | 0.903 | 0.746 |
| GT-Base | 0.822 | 0.768 | 0.837 | 0.832 | 0.834 |
| Bert-TRGCN | 0.333 | 0.415 | 0.306 | 0.275 | 0.311 |
| TRGCN | **0.894** | **0.857** | **0.889** | **0.923** | **0.900** |

**Table 4.**

*The results of comparative experiments on the Twitter16 dataset*

| Method | Acc | NR F1 | FR F1 | TR F1 | UR F1 |
|---|---|---|---|---|---|
| DTC | 0.473 | 0.254 | 0.080 | 0.190 | 0.482 |
| RFC | 0.585 | 0.752 | 0.415 | 0.547 | 0.563 |
| SVM-TS | 0.574 | 0.755 | 0.420 | 0.571 | 0.526 |
| SVM-HK | 0.511 | 0.648 | 0.434 | 0.473 | 0.451 |
| GRU-RNN | 0.633 | 0.617 | 0.715 | 0.577 | 0.527 |
| BU-RVNN | 0.718 | 0.723 | 0.712 | 0.779 | 0.659 |
| Rumor2vec | 0.852 | 0.857 | 0.769 | 0.927 | 0.850 |
| HDGCN | 0.865 | 0.820 | 0.863 | **0.930** | 0.863 |
| GCRES | 0.888 | 0.801 | 0.877 | 0.912 | 0.919 |
| GT-Base | 0.834 | 0.815 | 0.813 | 0.885 | 0.797 |
| Bert-TRGCN | 0.365 | 0.4774 | 0.295 | 0.143 | 0.325 |
| TRGCN | **0.901** | **0.870** | **0.939** | 0.872 | **0.905** |

Based on the experimental results, the comparison of baseline algorithms reveals a clear advantage for the deep learning approach over methods using manual features. Deep learning models effectively extract abstract features from rumor-related information, allowing them to capture effective features with high accuracy and achieve superior performance. This finding underscores the significant value and the urgent need for further investigation into advanced deep learning methods to identify misinformation.

Traditional deep learning methods, particularly those that do not leverage graph neural networks (GNNs), face significant theoretical limitations. These methods lack explicit mechanisms for modeling graph structures, making it difficult to effectively capture the topological dependencies in information propagation paths. As a result, they often suffer from systematic biases when attempting to characterize the dynamics of rumor propagation. Specifically, these approaches do not incorporate graph-theoretic operators to resolve multi-hop interactions between nodes, leading to sub-optimal performance in tasks such as predicting propagation influence and inferring traceability. In contrast, graph convolutional networks (GCNs) address these limitations by using spectral domain transformations and neighborhood aggregation to resolve multi-scale network structures. This framework not only explicitly encodes the structural similarities between nodes but also captures higher-order propagation patterns through deep stacked layers. Additionally, the message-passing mechanism in GCNs is theoretically equivalent to low-pass filtering the eigenspace of the Laplace matrix, which enables effective modeling of the cascading diffusion effect and community homogenization phenomena in rumor propagation.

By learning the graph representation end-to-end, the GCN framework can accurately quantify message propagation paths and predict potential probability distributions. This provides a robust mathematical foundation for constructing interpretable models of rumor propagation dynamics.

Lastly, our proposed TRGCN method outperforms all baseline models on both datasets, achieving accuracies of 89.3% and 90.1%, respectively. The model integrates the strengths of both GCNs and Transformers, combining graph-level dependencies, structural, and textual information to enhance representation learning. The GCN component identifies inter-graph dependencies and exploits the architecture of the misinformation spread network, facilitating efficient feature identification. Meanwhile, the Transformer model incorporates long-range nodes through positional encoding, which enhances the model's capacity to recognize long-range relationships and integrate contextual information. This combination of GCN and Transformer components enhances the model's discriminative power and generalization ability, ultimately leading to superior performance.

***Ablation Study***

Ablation studies provide a quantitative means of assessing the contribution of individual model components to rumor detection performance. To evaluate the importance of key modules and guide architectural design, we conduct incremental ablation experiments by systematically removing each component and measuring the resulting drop in performance. The specific configurations are as follows:

**NoMultiheadAttention:** The Multi-head attention mechanism is removed to evaluate its role in capturing diverse feature patterns across representation subspaces and in emphasizing critical components of the rumor propagation graph.

**NoPositionalEncoding:** Positional encoding is disabled to assess the importance of sequential and structural position information in modeling the propagation structure.

**NoResidual:** Residual connections in both the GCN and Transformer layers are removed to examine their contribution to stable training and performance in deep network configurations.

**GCNOnly:** Only the GCN model used in the study is retained, with all other components removed.

**TransformerOnly:** Only the Transformer module is used, with all other components removed.

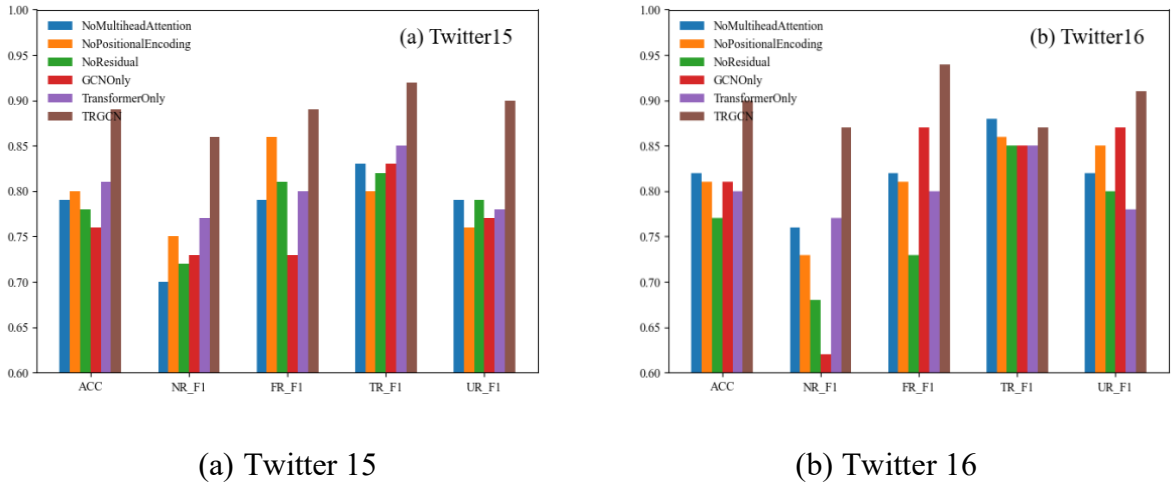

(a) Twitter 15           (b) Twitter 16

**Fig. 3 Experimental results for various component model variations.** Each cluster of bars represents a model variant, and the x-axis labels denote the evaluation metrics. The bars are color-coded to indicate the corresponding model.

As shown in **Fig. 3**, a series of experiments conducted on the different versions of our model's fundamental components showed that the complete TRGCN model consistently outperforms variants missing any of its key components. This outcome underscores the thoroughness of our proposed strategy and the significance of each distinct component.

**Conclusion and future work**

The persistent spread of rumors poses serious risks to social development, and the accelerated diffusion enabled by the internet has magnified the impact, weakened public trust, and endangered social stability. The emergence of artificial intelligence offers promising solutions for rumor detection. In this study, we propose the TRGCN model, which integrates two state-of-the-art deep learning architectures: Graph Convolutional Networks (GCNs) and Transformers, to capture multi-dimensional features of rumors from both content and propagation patterns. Specifically, GCNs are employed to model the structural information of rumor diffusion, while textual content is encoded using word embeddings. The positional encoding and Multi-head attention mechanism further enhance the model's capacity to jointly learn from content semantics, propagation dynamics, and sequential dependencies, thereby improving detection accuracy. We assess our model's robustness and performance using two open datasets, validating its real-world applicability.

The harm caused by rumor spreading should not be underestimated, making timely detection and governance essential in information dissemination. In subsequent research, we intend to extend the model to detect multimodal information, incorporating modalities such as images and videos to further enhance rumor detection. Additionally, we aim to investigate dynamic predictions of communication links and develop more efficient models to forecast the evolution of rumor trends, thereby improving the practical application of rumor classification.

**Data availability**

The code employed in this study is available on GitHub upon acceptance: https://github.com/Qingkongyan/TRGCN.git

## Funding

This work was supported by National Social Science Funds of China (Grant No. 22BSH025), Zhejiang Provincial Natural Science Foundation of China (No. LTGG24F030002), National Natural Science Foundation of China (Grant No. 61803047), the Social Sciences Fund of Jiangsu Province (Grant No. 24XWB004), the Jiangsu Qing Lan Project, and the Special Research Project on the Digital Transformation of Higher Education and the Practice of Educational Modernization in Jiangsu Province (Grant No. 2024CXJG061).

## Competing interests

The authors declare no competing interests.

## Ethical approval

This article does not contain any studies with human participants performed by any of the authors.

## Informed consent

This article does not contain any studies with human participants performed by any of the authors.

## Additional information

**Correspondence** and requests for materials should be addressed to D.Y., C. W., and K.S.

**Fig. 1 Example of a rumor propagation network.** Each node in the network represents an individual social media user who interacts with a particular post containing the rumor. The blue node indicates the root user, the original poster who first introduces the piece of information (often the source of the rumor). Red nodes represent refuters, users who actively respond to or challenge the rumor by posting corrections, counterarguments, or factual information aimed at debunking the false claim. Green nodes are rumor spreaders, individuals who share or forward the rumor without questioning its validity, thus contributing to its dissemination. The diagram illustrates how the rumor spreads among different types of users over time steps. Arrows denote the forwarding relationships between users and the post.

**Fig.2  The architecture of the TRGCN module.** TRGCN consists of three components: (1) rumor representation through the integration of textual content and propagation structure, followed by the integration of node features with positional encoding; (2) a multi-level fusion model that combines Graph Convolutional Networks (GCNs) and Transformer architectures, with their representations integrated through attention mechanism layers; and (3) the rumor classification task, which determines the veracity of a given piece of information (e.g., true, false, or unverified).

**Fig. 3  Experimental results for various component model variations.**  Each cluster of bars represents a model variant, and the x-axis labels denote the evaluation metrics. The bars are color-coded to indicate the corresponding model.